\documentstyle[12pt,a4]{article}
\begin{document}
\begin{flushright}
DFPD 92/TH/22
\end{flushright}
\begin{flushright}
SISSA--78/92/AP
\end{flushright}
\vskip 1cm
\setcounter{page}{0}
\centerline{\LARGE\bf Cosmic $\Delta B$ from Lepton Violating Interactions}
\vskip 0.6cm
\centerline{\LARGE\bf at the Electroweak Phase Transition}
\vskip 1.6cm
\centerline{Antonio Masiero$^{\dag}$ and Antonio Riotto$^{\S}$}
\vskip 1cm
\centerline{$^{\dag}$ \sl{Istituto Nazionale di Fisica Nucleare, Sezione di
Padova, 35100  Padova, Italy.}}
\vskip 0.5cm
\centerline{$^{\S}$ \sl{International School for Advanced Studies, SISSA}}
\vskip 0.15cm
\centerline{\sl{via Beirut 2-4, I--34014 Trieste, Italy.}}
\vskip 2.2 truecm
\centerline{\Large\bf Abstract}
\vskip 0.8cm
\hspace{0.3 cm} We propose a new mechanism for late cosmological baryon
asymmetry in models with first order electroweak phase transition. Lepton
asymmetry arises through the decay of particles produced out of equilibrium
in bubble collisions and is converted into baryon asymmetry by sphalerons.
Supersymmetric models with \underline{explicitly} broken $R$--parity may
provide a suitable
framework for the implementation of this mechanism.
\newpage
\hoffset=0.4cm
\voffset=-1 truecm
\normalsize
\def\simless{\mathbin{\lower 3pt\hbox{$\rlap{\raise 5pt\hbox{$\char'074$}}
\mathchar"7218$}}}
\def\simgreat{\mathbin{\lower 3pt\hbox{$\rlap{\raise 5pt \hbox{$\char'076$}}
\mathchar"7218$}}}
\def\\{\vskip 0cm}
\def\theequation{\arabic{equation}}
\setcounter{equation}{0}
\baselineskip 24pt
\hspace{0 cm}
The realization that the baryon $\left( B \right)$ and the lepton $\left( L
\right)$ violating quantum effects in the standard electroweak theory are
efficient
 at high temperature [1] has sparkled a widespread interest on the issue of
late cosmological $B$ asymmetry production. The problem is that any
matter--antimatter asymmetry created at some superheavy scale [2] can be
easily wiped out
by the $B$--violating quantum effects, which are abundantly in equilibrium
throughout
 all the history of the early Universe until its temperature drops down to
the Fermi scale
 range [1]. If this is indeed the case, then we are faced with the vital
problem
of originating a new cosmological $B$ asymmetry at the scale of the
Electro Weak  Phase Transition (EWPT).\\ A key ingredient for the development
of a sizeable late cosmological $B$ asymmetry is that the EPTW be of first
order. This seems necessary for the implementation of the so--called out of
equilibrium condition which, together with $B$ and $CP$ violation, constitutes
 one of the essential requirements to generate a net $\Delta B$ [3]. There
already
exist several studies of $\Delta B $ production at the electroweak scale in the
 Standard Model (SM) or in minimal extensions of it [4]. In SM the
$B$--violating quantum effects play
a twofold role [5]: on one hand they are responsible for the washing of any
preexisting $B$ asymmetry, but, on the other hand, they are invoked for the
production of the new $\Delta B$ at the Fermi scale. In general,
in order for these scenarios  at the electroweak scale to work, the
$B$--violating quantum effects must be
sufficiently suppressed after the accomplishment of the EWPT, or else they
can wipe
out the $B$ asymmetry once again. This condition translates into an upper
bound on the mass of the light neutral physical Higgs [5,6]. This limit is in
contrast with or dangerously close to the present LEP lower bound on Higgs
masses [7], depending on the details of the different analyses and whether the
SM [5] or the Minimal Supersymmetric Standard Model (MSSM) are considered
[6]. \\
In this paper we focus on a different situation, where at the Fermi scale some
\underline{new} $B$-- and/or $L$--violating phenomenon is effective in
addition to the $B$-- and $L$--violating quantum effects of the SM. In
particular, we consider the case in
in which new $L$--violating interactions are responsible for producing a
cosmological
$\Delta L$ at the EWPT, while the $B$-- and $L$--violating quantum effects are
subsequently called into play in order to transform this $\Delta L$ into the
cosmological $\Delta B$ \footnote{Baryogenesis via Leptogenesis was first
considered in [8], however, in our model, the lepton asymmetry and the out
of equilibrium condition are obtained in  completely different manners.}.
Notice
that in our approach the requirement concerning the quantum effects just after
the  of the EWPT is opposite to that previously imposed in SM and
MSSM: namely, they must be still efficient in order to convert $\Delta L$ into
$\Delta B$. Hence, the dangerous upper bound on the physical Higgs mass is
removed in
our case. We provide an illustration of these ideas in the context of
supersymmetric
(SUSY) models [9] where $R$--parity is \underline{explicitly} broken in the
leptonic sector [10]. Unless the violation is extremely tiny, the
$L$--violating transitions originated
 by $R$--parity breaking are in equilibrium down to temperatures at which the
EWPT
 occurs. We show that the $R$--violating interactions can originate an adequate
$\Delta L$ in theories with a first order EWPT and such $\Delta L$ can be
efficiently converted into $\Delta B$ by $B$--and $L$--violating quantum
effects
still operative at the end of the phase transition.\\
Whether or not the EWPT in SM is of first order is still an open issue much
debated in the present literature [11]. The new twist in this discussion is
represented by the inclusion of infrared effects which contribute
non--negligibly
to the definition of the order of EWPT. It seems that, in any case, the
transition
may remain first order even after their inclusion, although possibly weaker
than what appeared in previous analyses [12]. Even a weak first order EWPT can
constitute a suitable framework for our proposal.\\
We consider the one--loop effective potential coming
from the SUSY model whose Higgs sector contains, in addition to the usual two
Higgs doublet superfields  $\hat{H}_{1}^{0}$ and $\hat{H}_{2}^{0}$, one
or more singlet superfields $\hat{N}$'s [13], with a typical coupling in the
superpotential $\xi_{i} \hat{H}_{1}^{0}\hat{H}_{2}
^{0}\hat{N}_{i}$. The theory simplifies in the limit in which the SUSY
breaking mass
scale $\tilde{m}$ is much larger than the weak scale. Indeed, only one
combination
of the two neutral Higgs bosons $H_{1}^{0}$ and $H_{2}^{0}$ remains light. For
instance, in the case of one singlet $\hat{N}$
\begin{equation}
h=\cos\beta\; {\rm Re}H_{1}^{0} + \sin\beta\;{\rm Re}H_{2}^{0} +{\cal O}
\left(\frac{v}{x}\right){\rm Re} N,
\end{equation}
where
\begin{equation}
{\rm Re}\langle H_{1}^{0}\rangle\equiv v_{1},\;\; {\rm Re}\langle H_{2}^{0}
\rangle
\equiv v_{2},\;\; \tan\beta\equiv \frac{v_{2}}{v_{1}},
\end{equation}
and
\begin{equation}
{\rm Re} \langle N \rangle\equiv x,\;\; v\equiv\sqrt{v_{1}^{2}+v_{2}^{2}},
\end{equation}
Here we have assumed $x\gg v$.\\Hence, at energy scales lower than
$\tilde{m}$, the theory approximately
reduces to the standard model with only one Higgs doublet. The one--loop
potential at finite temperature $V\left(h,T\right)$ can be written as the
sum of
the classical potential, the Coleman--Weinberg quantum corrections [14] and
a temperature--dependent part [15]. Calling $T_{0}$ and $v\left(T_{0}\right)$
the temperature at which the
potential is flat at the origin and the vacuum expectation value of
$h$ at $T=T_{0}$, respectively, we obtain
\begin{equation}
\frac{v\left(T_{0}\right)}{T_{0}}=\frac{\alpha}{\lambda_{T}},
\end{equation}
where $\alpha$ and $\lambda_{T}$ are the coefficients of the
cubic and quartic term in $h$, respectively. $\lambda_{T}$ can be written as
$\lambda_{T}=\lambda -\frac{1}{2}K$, where $\lambda$ is the coefficient of
the $h^{4}$ term in SM and $K$ denotes the correction coming from
the supersymmetric contribution to the effective potential. Given our
assumption that $\tilde{m}\gg m_{W}$, we can expand the mass squared
of each SUSY particle in power of $\tilde{m}$ [6]
\begin{equation}
m_{i}^{2}=\tilde{m}^{2}+g_{i}h^{2}+{\cal
O}\left(\frac{1}{\tilde{m}^{2}}\right).
\end{equation}
Then $K$ is found to be [6]
\begin{equation}
K=\sum_{i}\pm \frac{N_{i} g_{i}^{3}}{8\pi^{2}}\frac{v^{2}}{\tilde{m}^{2}}.
\end{equation}
The sum in (6) extends over all bosons (+) and fermions ($-$) present in
the theory with number of degrees of freedom $N_{i}$.\\
A non zero value of $v\left(T_{0}\right)/T_{0}$ signals a first order phase
transition. Here comes the abovementioned crucial difference between
baryogenesis schemes
where $\Delta B$ is generated before the end of the EWPT and our scenario.
In the former case, in order to avoid the cosmic $\Delta B$
be washed out soon after the phase transition, a lower bound on
$v\left(T_{0}\right)/T_{0}$ must be imposed,
$v\left(T_{0}\right)/T_{0}\simgreat
1.3$ [5,16]. From eq. (4) it is easy to realize that this bound implies
an upper bound on the Higgs boson mass in SM where, roughly, $m_{h}$ is
proportional to $\lambda^{1/2}$. In our scenario the anomalous
electroweak processes must be still effective soon after the accomplishment of
the EWPT and hence, no upper bound on $m_{h}$ has to be imposed. However,
the presence of a singlet $\hat{N}$ is still relevant also in our treatment.
It was recently observed [17] that in SM the
EWPT may proceed via percolation of subcritical bubbles, originated by
thermal fluctuation, unless the Higgs boson mass, and consequently $\lambda$,
are small enough, $m_{h}\simless$ 100 GeV, according to the recent estimate
of Dine et al. [11]. Thus, if one decreases the value of
$\lambda_{T}$, the first order phase transition
is made easier. Taking the result for $K$ from ref. [6] for the case of
the SUSY model with two singlets $\hat{N}_{1}$ and $\hat{N}_{2}$, one
can see that it is indeed possible to lower $\lambda_{T}$ considerably with
respect to its value in SM \footnote{
In the presence of only one singlet $\hat{N}$, the fermionic contribution
increases $\lambda_{T}$. We have checked that the inclusion of the bosonic
contribution may decrease $\lambda_{T}$ enough only if one performs a
fine--tuning.},
 without
decreasing the value of $m_{h}$. In
fact, one must be careful not to decrease $\lambda_{T}$ too much, or else
$v\left(T_{0}\right)/T_{0}$ becomes too large and the anomalous electroweak
phenomena are no longer operative after the EWPT. Taking $m_{h}\approx
m_{t}\approx$ 100 GeV, a value of $\lambda_{T}\approx
2\times 10^{-2}$ generates a suitable phase transition for the baryogenesis
scenario.\\
We come now to a short description of the dynamics of bubble nucleation
and collision in a first order phase transition. At high temperatures,
$T\gg T_{0}$, the Universe is in the symmetric phase and there exists a unique
minimum of $V(h,T)$ at $\langle h\rangle=0$. As temperatures lowers, a second
minimum appears and becomes degenerate with the minimum at $\langle h\rangle
=0$ at [18]
\begin{equation}
T_{C}=T_{0}\left(1-\frac{2}{9}\frac{\alpha^{2}}{\omega\lambda_{T}}\right)
^{-1/2},
\end{equation}
where $\omega$ is the coefficient of the $h^{2}T^{2}$ term in the potential.\\
The Universe decays from its false vacuum state at $\langle h\rangle =0$
by bubble nucleation. Nucleation takes place at $T_{NUC}\sim T_{0}\sim$ 150
GeV, if $\lambda_{T}\approx 2\times 10^{-2}$. The difference between the
false and the true vacuum energy densities
at $T\sim T_{0}$ reads
\begin{equation}
\rho_{VAC}\approx \frac{4}{9}\frac{\alpha^{2}\omega}{\lambda_{T}^{2}}
T_{0}^{4},
\end{equation}
and is transformed into potential energy in the bubble walls whose energy
density is (see Enqvist et al. [18])
\begin{equation}
\eta\approx \frac{2^{3/2}}{3^{4}}\frac{\alpha^{3}}{\lambda_{T}^{5/2}}T_{0}^{3}.
\end{equation}
At $T<T_{NUC}$ bubbles keep growing. If they expand by a factor $\gamma$ since
$T_{NUC}$, their kinetic energy becomes, roughly speaking,
${\cal O}(\gamma)$ larger than the potential
energy [19]. The total energy of the bubble reads
\begin{equation}
E_{TOT}\approx 4\pi R^{2}\eta\gamma,
\end{equation}
where $R$ is the radius of the expanding bubble.\\
When bubbles collide the energy (10) can be released with two different
mechanisms: $i$) \underline{direct particle production} due to quantum
effects [20] and $ii)$ \underline{generation of coherent scalar waves} [19,20]
which decay into light particles (such a production appears to be less
useful for our
scenario). In both cases the resulting particles
are obviously produced with a distribution far from the equilibrium
distribution and this represents the crucial ingredient in our scenario to
implement the out of equilibrium condition necessary to originate a net
$\Delta L$.\\
Our aim now is to estimate the number of particles of each species
generated through the
abovementioned direct particle production mechanism. The mean energy of a
particle produced in the bubble collision is of order of the inverse
thickness of the wall, $\langle E\rangle \sim \Delta_{\gamma}^{-1}\equiv
\Delta^{-1}\gamma$, where $\Delta\approx 6\sqrt{2\lambda_{T}}/\left(\alpha
T_{0}\right)$ is the size of the wall when it is formed and we are assuming
that $\gamma\approx 1/\sqrt{1-v_{w}^{2}}$, $v_{w}$
being the wall velocity [19].\\
The number density of particles of species $\tilde{\chi}$ produced in the
collision
is of order
\begin{equation}
n_{\tilde{\chi}}\approx f_{\tilde{\chi}}\:\rho_{VAC}\:\Delta_{\gamma}
\end{equation}
where $f_{\tilde{\chi}}$ paramatrizes the fraction of particles of species
$\tilde{\chi}$
produced in the direct bubble collision. We proceed to give an estimate of
this factor $f_{\tilde{\chi}}$. Following ref. [20], one computes the
probability
of producing particles $\tilde{\chi}$'s in the decay of off--shell Higgs fields
describing the dynamics of the bubble collision. Such probability reads
\begin{equation}
{\cal P}\simeq \int \frac{d^{4}k}{\left(2\pi\right)^{4}}\left|{\cal
H}(k)\right|
^{2}\;{\cal F}(k^{2})\;\Theta\left(k^{2}-4m^{2}_{\tilde{\chi}}\right),
\end{equation}
where $k=(k_{0},k_{x},k_{y},k_{z})$, ${\cal H}(k)$ denotes the Fourier
transform of the suitable Higgs field
configuration, ${\cal F}(k^{2})$ is the imaginary part of the two points
one--particle irreducible Green function constructed from the
$h$-$\tilde{\chi}$-$\tilde{\chi}$ coupling and $m_{\tilde{\chi}}$ represents
the mass of the particle
$\tilde{\chi}$. The important message of (12) is that particles
$\tilde{\chi}$'s with
masses larger than $m_{h}/2$ can be produced through the decays of virtual
Higgs fields. A simple $ansatz$ for the field configuration describing
the two bubble collision is
\begin{equation}
h(t,z)\approx\left\{ \begin{array}{ll}
0 & \mbox{if $ v_{w}t<z<-v_{w}t,\;\;\;\;\;\;t<0$}\\
0 & \mbox{if $-v_{w}t<z<v_{w}t,\;\;\;\;\;\;t>0$}\\
v\left(T_{0}\right)& \mbox{otherwise,}
\end{array}\right.
\end{equation}
where we have approximated the two colliding
bubbles as plane--symmetric walls moving along the $z$ direction with velocity
$v_{w}$. The Fourier transform of (13) yields
\begin{equation}
{\cal H}(k)=\frac{4 v_{w} v\left(T_{0}\right)}{k_{0}^{2}-k_{z}^{2}v_{w}^{2}}.
\end{equation}
{}From (12) we can estimate the number of particles $\tilde{\chi}$'s produced
per unit area
\begin{equation}
\frac{{\cal N}}{A}\approx \int \frac{dk_{z}dk_{0}}{\left(2\pi\right)^{2}}\;
\left|{\cal H}(k_{0},k_{z})\right|^{2}\;{\cal F}(k_{0}^{2}-k_{z}^{2})\;
\Theta\left(k_{0}^{2}-k_{z}^{2}-4m_{\tilde{\chi}}^{2}\right).
\end{equation}
The computation simplifies assuming the step--function configuration (13)
and the scattering to be elastic [20]. These approximations give accurate
results for $k_{z}\simless\Delta_{\gamma}^{-1}$ and $k_{0}\simless \left(
\Delta_{\gamma}^{-1}v_{w}\right)$. Note that the slow power law--fall at large
$k_{0}^{2}-k_{z}^{2}v_{w}^{2}$ in expression (14) is due to the fact
that $h(t,z)$ is discontinuous. For \underline{more realistic thick walls},
${\cal H}(k_{0},k_{z})$ would cut off exponentially for $k_{z}\simgreat
\Delta_{\gamma}^{-1}$ and $k_{0}\simgreat\left(\Delta_{\gamma}v_{w}\right)
^{-1}$ [20].\\
If we consider a typical Yukawa coupling of the Higgs field to fermions
$\tilde{\chi}$'s, ${\cal L}=gh\bar{\tilde{\chi}}\tilde{\chi}$, from eq. (15)
we obtain
\begin{equation}
\frac{{\cal N}}{A}\approx g^{2}\;v\left(T_{0}\right)^{2}{\rm ln}
\left(\frac{\gamma \Delta^{-1}}{2m_{\tilde{\chi}}}\right).
\end{equation}
{}From the last expression we can argue that there is the possibility
of producing particles with mass approximately up to $\gamma\Delta^{-1}$.
Hence, the value of $\gamma$ becomes crucial in our discussion. For our
mechanism for $\Delta L$ production to work it is important
to be able to generate particles with mass larger than $m_{h}$ through
bubble collisions and, thus, we need a large value of $\gamma$.
The value of this parameter has been the focus of interest of a few recent
analyses. In ref. [21] it is claimed that bubbles can expand with $v_{w}$
close to the speed of light. However Dine et al. [11] have identified relevant
mechanisms wich may slow down $v_{w}$. Their result depend on microsopic
conditions for themalization processes in the vicinity of the wall and a more
complete analysis of the Boltzmann equation to determine the departure
from equilibrium in the particle phase--space density across the wall
is needed [21]. Moreover, even a small gradient in temperature across the
wall can be crucial for accelerating
the wall itself, see Enqvist et al. [18] and ref. [22].\\
The estimate of $f_{\tilde{\chi}}$ comes from the direct comparison of
${\cal N}/A$
given in (16) and that calculated from (10)
\begin{equation}
\frac{{\cal N}}{A}\approx \frac{E_{TOT}}{\langle E\rangle A}\:f_{\tilde{\chi}}
\approx \eta\:\Delta_{\gamma}\:f_{\tilde{\chi}}\approx v\left(T_{0}\right)^{2}
f_{\tilde{\chi}},
\end{equation}
where the last equality is readily obtained using eqs. (4) and (9). We obtain
\begin{equation}
f_{\tilde{\chi}}\approx g^{2}\:{\rm ln}\left(\frac{\gamma\Delta^{-1}}{2m_
{\tilde{\chi}}}\right).
\end{equation}
We have now all the factors entering the expression (11) for the number density
of particles $\tilde{\chi}$'s. Then, $\Delta L$  produced in $\tilde{\chi}$
decays is
\begin{equation}
\Delta L\equiv \frac{n_{l}-n_{\bar{l}}}{s}=\frac{n_{L}}{s}\approx
\frac{\varepsilon_{L}\,f_{\tilde{\chi}}\,
\rho_{VAC}\,
\Delta\,\gamma^{-1}}{\frac{2\pi^{2}}{45}g_{*\,S}\left(T_{0}\right)T_{0}^{3}},
\end{equation}
where
\begin{equation}
\varepsilon_{L}=\sum_{i}B_{i}\frac{\Gamma\left(\tilde{\chi}\rightarrow l_{i}
\right)-\Gamma\left(\bar{\tilde{\chi}}\rightarrow \bar{l}_{i}\right)}
{\Gamma\left(\tilde{\chi}\rightarrow all\right)},
\end{equation}
$B_{i}$ being the branching ratios of the decay mode $\tilde{\chi}\rightarrow
l_{i}$ and $g_{*\, S}\left(T_{0}\right)=106.75$ counts for the relativistic
degrees
of freedom contributing to $s$, the entropy density. In (19) we have taken
into account that the
"reheating" temperature is $\sim T_{0}$.\\
Using eq. (8) and taking $\alpha\sim 10^{-2}$, $\omega\sim 0.2$, and
$\lambda_{T}\sim 2\times 10^{-2}$, we obtain from eq. (19)
\begin{equation}
\Delta L\approx 10^{-2}\;\varepsilon_{L}\; f_{\tilde{\chi}}\;\gamma^{-1}.
\end{equation}
As we previously mentioned, we consider $R$--parity explicitly broken in the
leptonic sector [10] as the concrete framework for the illustration of
our ideas. The relevant $L$--violating terms in the superpotential are
\begin{equation}
W_{\Delta L\neq 0}=\frac{1}{2}\lambda_{ijk}\left[\hat{L}_{i},\hat{L}_{j}\right]
\hat{e}_{k}^{c}+
\lambda_{ijk}^{\prime}\hat{L}_{i}\hat{Q}_{j}\hat{d}_{k}^{c},
\end{equation}
where $i, j, k$ are generation indices, $\hat{L}$ and $\hat{Q}$ are the
left--handed lepton and quark doublet superfields, and $\hat{e}^{c}$ and
$\hat{d}^{c}$ denote (the charge conjugate of) the right--handed lepton and
charge $-1/3$ quark singlet superfields.\\
There already exist several experimental constraints on
the $\lambda$ and $\lambda^{\prime}$ couplings in (22) [23]. The strongest
bounds by far are those obtained from the cosmological considerations on the
 survival of the cosmic $\Delta B$ [24]. If we ask for the $L$--violating
interactions due to the couplings (22) to be constantly out of equilibrium
until the weak scale, so that they cannot wash out any $(B-L)$ asymmetry
previously generated, a severe upper limit
$\lambda, \lambda^{\prime}\simless 10^{-7}$, independently from the generation
indices is obtained\footnote{This is true only if one allows for the violation
of all partial lepton number symmetries in (22). It was pointed out that one
can evade these bounds if one of the lepton numbers $L_{e}, L_{\mu},
L_{\tau}$, say $L_{e}$, is preserved in (22), so that $\frac{1}{3}B-L_{e}$
is not washed out [25].}. \underline{In our case these limits are no
longer valid}
since we ask for the $L$--violating interactions in (22) to be still operative
at the weak scale in order to give rise to the late $\Delta L$. Other
phenomenological and
astophysical limits [26] exist, though they are substantially less tight, and
we shall
take them into account.\\
The scenarios we have in mind for the generation of $\Delta L$ goes as follows.
We assume the lightest supersymmetric particle (LSP) to be the lightest
neutralino $\tilde{\chi}$. Differently from what occurs in MSSM, the LSP can
now decay
through the $R$--violating couplings in (22). Since $\tilde{\chi}$ possesses
decay
channels with different $L$-- numbers, a net $\Delta L$ can develop in its
decay
if $CP$ is violated and the out of equilibrium condition is implemented.
$CP$ violation can be easily provided by the complexity of the $\lambda$ and
$\lambda^{\prime}$ couplings in (22). This is a remarkable advantage with
respect to late $\Delta B$ production in SM where $CP$ violation provided by
the CKM phase seems to be too small. Concerning the out of equilibrium
requirement,
this is automatically satisfied if the $\tilde{\chi}$'s are produced in the
bubble collision through their couplings to $h$.\\
$\tilde{\chi}$ is generally given by the superposition of the neutralinos of
the
model (notice, in particular, the presence of the singlet fermion $\tilde{N}$,
in the case of one singlet superfield $\hat{N}$)
\begin{equation}
\tilde{\chi}=Z_{11}\tilde{W}_{3}+Z_{12}\tilde{B}+Z_{13}\tilde{H}_{1}^{0}
+Z_{14}\tilde{H}_{2}^{0}+Z_{15}\tilde{N}.
\end{equation}
As shown in ref. [27], when the LSP is $\tilde{\chi}\approx \frac{1}{2}
\left(\tilde{H}^{0}_{1}+
\tilde{H}_{2}^{0}\right)-\frac{1}{\sqrt{2}}\tilde{N}$, it can be very massive,
$m_{\tilde{\chi}}={\cal O}(500)$ GeV. In this limit, the coupling of
$\tilde{\chi}$ to
$h$ reads
\begin{equation}
{\cal L}=Z_{15}\left(Z_{14}\,\cos\beta+Z_{13}\,\sin\beta\right)
\,h\, \bar{\tilde{\chi}}\frac{1-\gamma_{5}}{2}\tilde{\chi} +\mbox {h.c.}.
\end{equation}
The coefficient $g$ in eq. (18) is then
\begin{equation}
g=\sqrt{2}Z_{15}\left(Z_{14}\,\cos\beta +Z_{13}\sin\beta\right),
\end{equation}
where $\sqrt{2}$ takes into account that $\tilde{\chi}$ is a Majorana
fermion.\\
The dominant decay channel of the LSP is expected to be that of fig. 1
\footnote{If the $R$--couplings are nonvanishing, LSP can decay also into two
body final states via one--loop
diagrams, for instance into $W l_{i}$, but these channels are suppressed
by a factor ${\cal O}$(100) in comparison to that of fig. 1., see Campbell
et al. [24].}, whose
thermally averaged partial width is
\begin{equation}
\left\langle \Gamma\left(\tilde{\chi}\rightarrow t\,l_{i}\,d^{c}_{k}\right)
\right\rangle=
\Gamma\left(\tilde{\chi}\rightarrow t\,l_{i}\,d^{c}_{k}\right)\frac{K_{1}\left(
m_{\tilde{\chi}}/T_{0}\right)}{K_{2}\left(m_{\tilde{\chi}}/T_{0}\right)},
\end{equation}
where $K_{n}$ denotes the modified Bessel functions and the prefactor
in front of the usual (zero--temperature) decay rate can be interpreted as a
time dilation factor. A straightforward calculation leads to
\begin{eqnarray}
\left\langle \Gamma\left(\tilde{\chi}\rightarrow t\,l_{i}\,d_{k}^{c}\right)
\right
\rangle
&\approx&\frac{\alpha_{W}}{16}\left(\frac{m_{t}}{m_{W}}\right)^{2}\frac{1}{
768\pi^{3}}\frac{m_{\tilde{\chi}}^{5}}{\left(m_{\tilde{t}_{L}}^{2}+T_{0}^{2}
\right)
^{2}}\left|\lambda_{13k}^{\prime}\right|^{2}\times\nonumber\\
&\times&\left(1-\frac{3}{2}\frac{T_{0}}{m_{\tilde{\chi}}}+\frac{15}{8}\frac{
T_{0}^{2}}{m_{\tilde{\chi}}^{2}}\right)Z_{14}^{2},
\end{eqnarray}
where, in our case, $Z_{14}=\frac{1}{2}$.\\
The expansion rate of the Universe is given by $H\simeq 17 T^{2}/M_{P}$,
where $M_{P}\simeq1.2\times 10^{19}$ GeV is the Planck mass.
Neutralinos can decay after their production if, roughly speaking,
$\Gamma\simgreat H$, which translates into the bound
\begin{equation}
\left|\lambda^{\prime}_{13k}\right|\simgreat 9.2\times 10^{-5}
\left(\frac{500\;\mbox {GeV}}{m_{\tilde{\chi}}}\right)^{5/2}\left(\frac{T_{0}}{
150\;\mbox {GeV}}\right)\left(\frac{m_{\tilde{t}_{L}}}{1\;\mbox {TeV}}\right)
^{2}\left(\frac{100\;\mbox {GeV}}{m_{t}}\right).
\end{equation}
The interference of the diagrams of figs. 1 and 2 gives rise to a non
vanishing $\varepsilon_{L}$ of order
\begin{equation}
\varepsilon_{L}\simeq \frac{1}{16\pi}\frac{\sum_{ikmst}\left(\lambda^{\prime *}
_{itm}
\lambda_{stm}^{\prime}\lambda_{s3k}^{\prime *}\lambda_{i3k}^{\prime}
\right)}{\sum_{ik}\left|\lambda_{i3k}^{\prime}\right|^{2}}.
\end{equation}
Actually, this expression must be multiplied by a suppression factor which
takes into account that part of the neutralinos produced out of equilibrium by
our mechanism may thermalize before decaying, e.g. through
the $\tilde{\chi}\,t\rightarrow
\tilde{\chi}\,t$ scattering, and, consequently, they cannot give contribution
to
$\varepsilon_{L}$. A rough estimate for such a suppression factor is
given by $\left\langle\Gamma\left(\tilde{\chi}\rightarrow t\,l_{i}\,
d_{k}^{c}\right)\right\rangle/\left\langle\Gamma\left(t\,\tilde{\chi}
\rightarrow t\,
\tilde{\chi}\right)\right\rangle\approx 10^{3}\left|\lambda^{\prime}_{13k}
\right|^{2}$.\\
 In order to prevent a washing out of the lepton asymmetry, we impose that $
\Delta L=1$ scatterings induced by $\lambda$ and/or $\lambda^{\prime}$
couplings are out of equilibrium at $T_{0}$. The predominant process is
shown in fig. 3. Since $T_{0}<m_{\tilde{\chi}}$, this process has a thermally
averaged rate
\newpage
\begin{eqnarray}
\Gamma_{\Delta L=1}=n_{EQ}(T)\left\langle \sigma v\right\rangle&\simeq&
\frac{27 Z_{14}^{2}}{32\pi^{3}\zeta(3)}\left|\lambda_{i3k}^{\prime}
\right|^{2}\frac{T_{0}^{3}m_{\tilde{\chi}}^{2}}{256}\alpha_{W}\times\nonumber\\
&\times&\left(\frac{m_{t}}{m_{W}}\right)^{2}\frac{1}
{\left(m_{\tilde{t}_{L}}^{2}+T_{0}^{2}\right)^{2}}{\rm e}^{-m_{\tilde{\chi}}
/T_{0}},
\end{eqnarray}
where $\zeta(3)\simeq 1.202$ is the Riemann zeta function and $n_{EQ}$
denotes the equilibrium number density. Requiring
$\Gamma_{\Delta L=1}\simless H$ imposes
\begin{equation}
\left|\lambda_{i3k}^{\prime}\right|\simless 4.2\times 10^{-4}
\left(\frac{m_{\tilde{t}_{L}}}{1\;\mbox {TeV}}\right)^{2}
\left(\frac{500\;\mbox {GeV}}{m_{\tilde{\chi}}}\right)
\left(\frac{150\;\mbox {GeV}}{T_{0}}\right)^{1/2}\left(\frac{100\;\mbox {GeV}}{
m_{t}}\right).
\end{equation}
A similar upper bound is found requiring that inverse decays are out
of equilibrium. The products of neutralino decays will be brought to thermal
equilibrium
at rates of order $T_{0}$. At this temperature, anomalous $(B+L)$--violating
processes are in equilibrium, since $\Gamma_{sphaleron}\simeq \alpha_{W}^{4}
T{\rm e}^{-\frac{4\pi}{g_{W}}\frac{v\left(T_{0}\right)}{T_{0}}}\simgreat H$ for
$\lambda_{T}\approx 2\times 10^{-2}$ and the lepton asymmetry is converted
into a
baryon asymmetry $\Delta B=n_{B}/s$, whose present value is
($T_{tod}\simeq 2.7  ^{0}$K) [28]
\begin{equation}
\Delta B\simeq -\frac{21}{61}\frac{g_{*\,S}\left(T_{tod}\right)}{g_{*\,S}
\left(T_{0}\right)}\;\Delta L,
\end{equation}
where $g_{*\,S}\left(T_{tod}\right)/g_{*\,S}\left(T_{0}\right)$ takes
into account the increase of entropy since the EWPT up to now.\\
{}From eqs. (21), (29) and (32), we argue that a baryon asymmetry $4\times
10^{-11}\simless\Delta B\simless 5.7\times 10^{-11}$ [29] can be produced
for resonable values of the $R$--violating
couplings. For instance, taking $m_{\tilde{t}_{L}}\approx 5$ TeV, $\gamma
\approx 10^{2}$, $\lambda_{T}
\approx 2\times 10^{-2}$ and all the $\lambda^{\prime}$ couplings of the same
order, we
obtain $\lambda^{\prime}\simeq 8\times 10^{-3}$. Note that this value is
in agreement with the upper bounds given in [26].\\
Notice that our scheme makes use of the $R$--parity violating couplings
in the leptonic sector. Previous proposals to use $R$--parity violation
for late $\Delta B$ production invoked $R$--breaking in the baryonic
sector within contexts quite different to implement the out of
equilibrium condition [30].\\
In conclusion, we have proposed a scheme for late baryon asymmetry in a first
order EWPT through the interplay of lepton violating interactions
and anomalous $B$-- and $L$--violating quantum effects: $i$) $\Delta L$ is
generated in the decay of the LSP produced out of equilibrium
in bubble collision and $ii$) $\Delta L$ is converted into $\Delta B$ by
sphalerons still operative after the EWPT. Obviously, a thorough discussion
of several open issues of EWPT is eagerly awaited for to provide a more
comprehensive and detailed scheme for late $\Delta B$ production along
the lines suggested in this paper.
\vskip 1cm
\centerline{\Large\bf Acknowledgements}
\vskip 0.5cm
\hspace{0 cm}
We should like to thank R. Barbieri, K. Enqvist, G. Giudice, E.W. Kolb,
S. Matarrese,
J. Miller, O. Pantano, S. Petcov and F. Zwirner for many helpful and
stimulating
discussions.
\newpage
\begin{description}
\centerline{{\LARGE\bf References}}
\setcounter{page}{14}
\hoffset=0.4cm
\voffset=-1 truecm
\vspace{1.5cm}
\item[[1]] V.A. Kuzmin, V.A. Rubakov and M.E. Shaposhnikov, Phys. Lett.
{\bf B155} (1985) 36; see also the review by V.A. Matveev, V.A. Rubakov,
A.N. Tavkhelidze and M.E. Shaposhinokov, Uspeckhi Fiz. Nauk {\bf 156} (1988)
253.
\item[[2]] For a review see, for instance, A.D. Dolgov and Ya.B. Zel'dovich,
Rev. Mod. Phys. {\bf 53} (1981) 1; E.W. Kolb and M.S. Turner, Mod. Phys. Lett.
{\bf A2} (1987) 285.
\item[[3]] A.D. Sakharov, Pis'ma Zh. Eksp. Teor. Fiz. {\bf 5} (1967) 32
and JETP Lett. {\bf 5} (1967) 24; M. Yoshimura, Phys. Rev. Lett. {\bf 41}
(1978) 28; S. Dimopoulos and L. Susskind, Phys. Rev. Lett.
{\bf D18} (1978) 4500; D. Toussaint, S.B. Tremain, F. Wilczek and A. Zee,
Phys. Rev. {\bf D19} (1979) 1036; S. Weiberg, Phys. Rev. Lett. {\bf 42}
(1979) 850; D.V. Nanopoulos and S. Weinberg, Phys. Rev. {\bf D20} (1979)
2484.
\item[[4]] For a review of Electroweak Baryogenesis, see A.D. Dolgov,
YITP/K 90 preprint (1991), submitted to Phys. Rep..
\item[[5]] A.I. Bocharev and M.E. Shaposhnikov, Mod. Phys. Lett. {\bf A2}
(1987) 417; A.I. Bocharev, S.V. Kuzmin and M.E. Shaposhnikov, Phys. Lett.
{\bf B244} (1990) 275; M.E. Shaposhnikov, CERN preprint TH-6319/91 (1991).
\item[[6]] G.F. Giudice, UTTG-35-91 preprint (1991).
\item[[7]] M. Davier in {\sl Proceedings of the Lepton--Photon Symposium
and Europhysics Conference on High Energy Physics}, S. Hagerty, K. Petter
and E. Quercigh (eds.), Geneva, 1991.
\item[[8]] M. Fukugita and T. Yanagida, Phys. Lett. {\bf B174} (1986) 45;
M. Luty, Phys. Rev. {\bf D45} (1992) 455.
\item[[9]] For a review, see H.P. Nilles, Phys. Rep. {\bf 110} (1984) 1;
H. Haber and G. Kane, Phys. Rep. {\bf 117} (1985) 1.
\item[[10]] C.S. Aulakh and R.N. Mohapatra, Phys. Rev.
{\bf D119} (1983) 136; L.J. Hall and M. Suzuki, Nuc. Phys. {\bf B231} (1984)
419;
I.H. Lee, Nuc. Phys. {\bf B246} (1984) 120;
J. Ellis et al., Phys. Lett. {\bf 150} (1985) 142;
G.G. Ross and J.W. Valle, Phys. Lett. {\bf B151} (1985) 375;
S. Dawson, Nuc. Phys. {\bf B261} (1985) 297;
R.N. Mohapatra, Phys. Rev. {\bf D11} (1986) 3457;
E. Ma and P. Roy, Phys. Rev. {\bf D41} (1990) 988;
E. Ma and D. Ng, Phys. Rev. {\bf D41} (1990) 1005;
S. Dimopoulos, R. Esmailzadeh, L.J. Hall, J.P. Merlo and G.S. Starkmen,
Phys. Rev. {\bf D41} (1990) 2099; H. Dreiner and R.J.N. Phillips,
Nuc. Phys. {\bf B367} (1991) 591; L.E. Ib\`{a}\~{n}ez and G.G. Ross,
Nuc. Phys. {\bf B368} (1992) 3.
\item[[11]] D. Brham and S. Hsu, Caltech preprints CALT-68-1705 and
CALT-68-1762
(1991); M.E. Carrington, Univ. of Minnesota preprint TPI-MINN-91/48-T (1991);
M.E. Shaposhnikov, CERN Preprint TH/6319/91 (1991); M. Dine, R.G. Leigh, P.
Huet, A. Linde and D. Linde, SLAC-PUB 5741
preprint (1992).
\item[[12]] For a review, see M. Sher, Phys. Rep. {\bf 179} (1989) 273 and
references therein.
\item[[13]] P. Fayet, Nuc. Phys. {\bf B90} (1975) 104; R.K. Kaul and P.
Mijudmar,
Nuc. Phys. {\bf B199} (1982) 36; R. Barbieri, S. Ferrara and C.A. Savoy,
Phys. Lett. {\bf B119} (1982) 343; H.P. Nilles, M. Srednicki and D. Wyler,
Phys. lett. {\bf B120} (1983) 346;
J.M. Frere, D.R.T. Jones and S. Raby, Nuc. Phys. {\bf B222} (1983) 11;
J.P. Derendinger and C.A. Savoy, Nuc. Phys. {\bf B237} (1984) 307;
J. Ellis, J.F. Gunion, H.E. Haber, L. Roszkowski and F. Zwirner, Phys. Rev.
{\bf D39} (1989) 844.
\item[[14]] S. Coleman and E. Weinberg, Phys. Rev. {\bf D7} (1973) 1888.
\item[[15]] L. Dolan and R. Jackiw, Phys. Rev. {\bf D9} (1974) 3320;
S. Weinberg, Phys. Rev. {\bf D9} (1974) 3357.
\item[[16]] M. Dine, P. Huet and R.S. Singleton Jr., Santa Cruz preprint
SCIPP-91-08 (1991).
\item[[17]] M. Gleiser, E.W. Kolb and R. Watkins, Nuc. Phys. {\bf B364}
(1991) 411; M. Gleiser and E.W. Kolb, FNAL preprint FERMILAB-PUB-91/305-A;
N. Tetradis, DESY preprint 91-151 (1991).
\item[[18]] K. Enqvist, J. Ignatius, K. Kajantie, K. Rummukainen,
preprint HU-TFT-91-35 (1991); G.W. Anderson and LJ. Hall, LBL-31169 preprint
(1991).
\item[[19]] S.W. Hawking, I.G. Moss and J.M. Stewart, Phys. Rev. {\bf D26}
(1982) 2681.
\item[[20]] R. Watkins and L.M. Widrow, preprint FERMILAB-Pub-91/164-A (1991).
\item[[21]] N. Turok, Phys. Rev. Lett. {\bf 68} (1992) 1803.
\item[[22]] O. Pantano and A. Riotto, in preparation.
\item[[23]] V. Barger, G.F. Giudice and T.Y. Han, Phys. Rev. {\bf D40} (1990)
1705.
\item[[24]] B. Campbell, S. Davidson, J. Ellis and K.A. Olive, Phys. Lett.
{\bf B256} (1991) 457 and CERN-TH-6208/91 preprint; W. Fischler, G.F. Giudice,
R.G. Leigh and S. Paban, Phys. lett. {\bf B258} (1991) 45.
\item[[25]] H. Dreiner, talk given at CERN, Geneva (1991).
\item[[26]] K. Enqvist, A. Masiero and A. Riotto, Nuc. Phys. {\bf B373}
(1992) 95.
\item[[27]] R. Flores, K.A. Olive and P. Thomas, Phys. Lett. {\bf B245}
(1990) 509.
\item[[28]] J.S. Harvey and M.S. Turner, Phys. Rev. {\bf D42} (1990) 3344.
\item[[29]] T.P. Walker, G. Steigman, D.N. Schramm, K.A. Olive and H.S. Kang,
Astrophys. J. {\bf 281} (1991) 51.
\item[[30]] S. Dimopoulos and L.J. Hall, Phys. Lett. {\bf B196} (1987) 135;
J. Cline and S. Raby, Phys. Rev. {\bf D43} (1991) 1781; R.J. Scherrer,
J. Cline,
S. Raby and D. Seckel, Phys. Rev. {\bf D44} (1991) 3760;
S. Mollerach and E. Roulet, preprint FERMILAB-PUB-91/340-A preprint (1991).
\end{description}
\begin{description}
\vspace{1.5cm}
\centerline{{\LARGE\bf Figure Captions}}
\vspace{1.5cm}
\item[Figure 1.] The tree--level Feynman diagram for the decay $\tilde{\chi}
\rightarrow t\,l_{i}\,d_{k}^{c}$ induced by the $\lambda^{\prime}_{i3k}$
couplings.
\item[Figure 2.] The one--loop Feynman diagram for the decay $\tilde{\chi}
\rightarrow t\,l_{i}\,d_{k}^{c}$, whose interference with that of Figure 1
gives rise to a non vanishing lepton asymmetry.
\item[Figure 3.] The Feynman diagram for the $\Delta L=1$ scattering
$l_{i}\,d_{k}^{c}\rightarrow t\tilde{\chi}$, which can wash out any lepton
asymmetry.
\end{description}
\end{document}